\documentclass[5p]{elsarticle}
\usepackage[utf8]{inputenc}
\usepackage{amsmath}
\usepackage{graphicx}
\usepackage[labelfont=bf]{caption}
\usepackage{wrapfig}
\usepackage{multicol}
\usepackage{floatrow}
\usepackage{xcolor}
\usepackage{siunitx}
\usepackage{gensymb}
\usepackage{textcomp}
\usepackage{color}
\usepackage{indentfirst}
\usepackage[colorinlistoftodos]{todonotes}
\usepackage{booktabs}
\usepackage{tabularx}
\newcolumntype{P}[1]{>{\centering\arraybackslash}p{#1}}
\let\vec\mathbf
\usepackage[breaklinks=true,colorlinks=true,linkcolor=blue,urlcolor=blue,citecolor=blue]{hyperref}
\biboptions{sort&compress}
\usepackage{cleveref}
\usepackage{float}
\usepackage{setspace}
\usepackage{soul}

\makeatletter
\def\ps@pprintTitle{
 \let\@oddhead\@empty
 \let\@evenhead\@empty
 \def\@oddfoot{}
 \let\@evenfoot\@oddfoot}
\makeatother

\newcommand{\pdv}[2]{\frac{\partial#1}{\partial#2}}

\begin{document}

\begin{frontmatter}

\title{Neural net modeling of equilibria in NSTX-U}

\author[Princeton]{J.T. Wai}
\ead{jwai@princeton.edu}

\author[PPPL]{M.D. Boyer}
\author[Princeton,PPPL]{E. Kolemen\corref{cor1}}
\ead{ekolemen@princeton.edu}

\cortext[cor1]{Corresponding author}

\address[Princeton]{Princeton University, Princeton, New Jersey, USA}
\address[PPPL]{Princeton Plasma Physics Laboratory, Princeton, New Jersey, USA}

\begin{abstract}

\indent 
Neural networks (NNs) offer a path towards synthesizing and interpreting data on faster timescales than traditional physics-informed computational models. In this work we develop two neural networks relevant to equilibrium and shape control modeling, which are part of a suite of tools being developed for the National Spherical Torus Experiment-Upgrade (NSTX-U) for fast prediction, optimization, and visualization of plasma scenarios. The networks include \textit{Eqnet}, a free-boundary equilibrium solver trained on the EFIT01 reconstruction algorithm, and \textit{Pertnet}, which is trained on the Gspert code and predicts the non-rigid plasma response, a nonlinear term that arises in shape control modeling. The NNs are trained with different combinations of inputs and outputs in order to offer flexibility in use cases. In particular, Eqnet can use magnetic diagnostics as inputs and act as an EFIT-like reconstruction algorithm, or, by using pressure and current profile information the NN can act as a forward Grad-Shafranov equilibrium solver. This forward-mode version is envisioned to be implemented in the suite of tools for simulation of plasma scenarios. The reconstruction-mode version gives some performance improvements compared to the online reconstruction code real-time EFIT (RTEFIT), especially when vessel eddy currents are significant. We report strong performance for all NNs indicating that the models could reliably be used within closed-loop simulations or other applications. Some limitations are discussed. 

\end{abstract}


\end{frontmatter}

{\let\thefootnote\relax\footnote{Code is available \url{github.com/PlasmaControl/nstxu-nns/}}}

\section{Introduction} \label{sec:intro}

In tokamaks, a critical aspect of achieving high performance plasma operation and maximizing scientific exploration is the development of fast and reliable model predictions of the plasma behavior. Fast model predictions are often a prerequisite to implementing real-time control algorithms. Additionally, if these are not available in real-time, but on a slightly slower timescale, they can be used by physics operators to inform decisions between shots. Machine learning is a tool that is gaining increasing usage in the nuclear fusion community and can be useful in this type of situation where it is desirable to quickly synthesize data into interpretable outputs. Some examples include its usage in profile predictions \cite{Jalalvand2021, Abbate2021, Boyer2021}, turbulent transport \cite{LI_2021}, identifying scalings laws \cite{Gaudio_2014, Murari_2010}, and predicting instabilities and disruptions \cite{Rea_2019, Lungaroni_2018, Montes_2019, Fu_2020, Pau_2019, Kates_Harbeck_2019, Piccione_2020}. 

In this work, we present two neural nets (NNs) \textit{Eqnet} and \textit{Pertnet} developed for the NSTX-U tokamak which are relevant to equilibrium and shape control modeling. Along with recent modeling of neutral beam heating \cite{Boyer2019} and electron profiles \cite{Boyer2021}, these neural nets are part of a suite of tools being developed for fast prediction, optimization, and visualization of plasma scenarios. Closed-loop simulations of a plasma discharge are generally too slow for usage between shots due to time-intensive steps of repeatedly solving the free-boundary Grad-Shafranov equation and linearizing around the given equilibrium. The neural nets here were developed to perform these two time-intensive tasks as a step towards fast simulation and eventually, numerical optimization of scenarios. Additionally, we explore and report on additional modes of operation of the networks that have alternative applications. 

\begin{figure}[H]
	\centering 
	\includegraphics[width = 0.8\columnwidth]{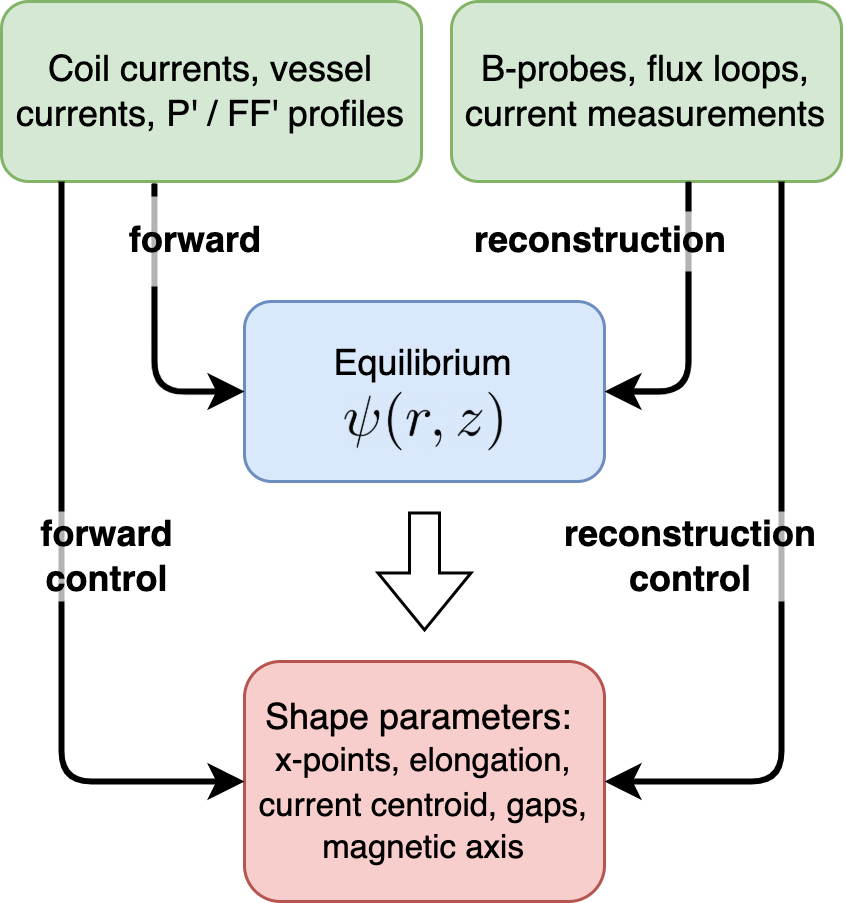}
	\caption{Diagram of the data inputs and outputs that were used to train different versions of Eqnet. The forward modes solve \cref{eq:GS} directly whereas the reconstruction modes are a diagnostic best-fit solution. The control modes predict extracted shaping parameters instead of the flux surfaces.}
	\label{fig:eqnet_modes}
\end{figure}

\Cref{sec:eqnet} introduces Eqnet, a NN-based architecture trained to predict the free-boundary plasma equilibrium based on various sets of inputs. In this work, we distinguish between several modes of operation (forward, forward-control, reconstruction, reconstruction-control) as diagrammed in \cref{fig:eqnet_modes}. The \textit{forward} mode predicts the equilibrium flux surfaces using the coil currents and plasma profiles as inputs. In contrast the \textit{forward-control} mode directly predicts control-relevant shaping parameters that are normally derived from the flux surfaces, including quantities such as the x-point locations, elongation, and outer gap distance. Similarly the \textit{reconstruction} and \textit{reconstruction-control} modes predict the same outputs, but use diagnostic information such as magnetic probes as inputs to the NN. Previous neural net equilibrium solvers \cite{Coccorese1994, Lister1991, Albanese1996, Bishop1995, LAGIN1993, Zhu2019, Prokhorov2020, Wang2015} were developed in the spirit of reconstruction-control mode--mapping diagnostics to shaping parameters--with some exceptions such as \cite{Joung2019} which reconstructs flux surfaces and \cite{vanMilligen1995} which is a fixed-boundary solver.

For control simulations, another important calculation is `linearizing the equilibrium' which includes obtaining the non-rigid plasma response, how current in the plasma shifts and redistributes in response to applied fields from the poloidal field (PF) coils. It is also common to model the plasma response to the normalized internal inductance $l_i$ and plasma poloidal beta $\beta_p$. This effect is nonlinear, equilibrium-dependent, and is addressed by codes such as Gspert/Gsupdate \cite{Welander2005, Welander2019} and Create-L/NL \cite{Albanese_1998, Albanese_2015}. In section \ref{sec:pertnet} we introduce Pertnet, a NN-based architecture for identifying the plasma response that was trained on the Gspert code. 

Eqnet and Pertnet were developed for the simulation goals stated above, although they have been trained to be flexible in inputs and outputs in order to demonstrate potential relevance to a variety of tasks. (To be clear, it is not a single model where inputs and outputs are toggled on and off, but multiple versions of the model trained with different inputs and outputs.) More specifically, some of the motivating applications are:

\begin{itemize}
    \item A fast plasma simulator for between shot scenario design (Eqnet and Pertnet): Presently available plasma ``flight simulators'' are not fast enough to run between shots, not to mention performing numerical optimization over the simulations. The key tasks in a flight simulator are solving for equilibria, and evolving currents and profiles according to the shape control model and plasma transport. Future work will focus on integrating Eqnet and Pertnet with NN-based profile evolution \cite{Boyer2021} in order to develop this simulator. 
    
    \item Increased availability and reliability of during-shot equlibrium reconstruction (Eqnet): NSTX-U relies on real-time EFIT (RTEFIT) for shape control measurements, but RTEFIT is not available until roughly 100-150ms into the shot when the plasma current is sufficiently high. A work-around method uses an additional gap controller \cite{Boyer2018} during this early time, but this adds complexity to the system in tuning the additional controller and dealing with controller transfer. A simpler method would be to use Eqnet as a backup for when RTEFIT is unavailable, since the reconstruction mode of Eqnet can use the same inputs and outputs as RTEFIT. Additionally, RTEFIT suffers inaccuracies when vessel currents are large or dynamic plasma changes lead to insufficient convergence time, which can be improved by using Eqnet (see \cref{sec:RTEFIT_comparison}). 
    
    \item Vertical growth rate monitoring (Pertnet): One of the outputs that can be obtained from Pertnet is the growth rate of the plasma vertical instability. Due to hardware actuator constraints, any individual tokamak has a limit on the growth rate that can be controlled before the plasma is lost to a wall collision, which was a difficulty for NSTX-U since it is a spherical tokamak with highly elongated plasmas. One possibility is to use Pertnet to monitor the growth rate in real-time so that corrective actions can be taken before an instability limit is reached. 
    
    \item Online updates to the shape controller (Pertnet): In most current tokamak shape control algorithms, feedback gains are calculated offline based on a single reference linearization, and these parameters are used for the entire discharge and across discharges. Using Pertnet, the equilibrium-dependent response predictions could be served to the controller rapidly ($<$1ms) during the shot to optimize controller performance. 
\end{itemize}

Development of these tasks will be the subject of future work. The rest of this paper is divided as follows. \Cref{sec:eqnet} introduces Eqnet including the data structure, model archictecture, prediction results, and a simple application problem of using Eqnet to design a novel equilibrium. \Cref{sec:pertnet} describes Pertnet, including the nonrigid plasma response model and relevant background, NN architecture, and performance results. 

\section{Eqnet}\label{sec:eqnet}

\subsection{Problem statement}
The goal of this neural net is to identify the plasma equilibrium $\psi(R,Z)$ which is the solution to the free-boundary Grad-Shafranov equation \cite{Grad1958, Shafranov1966}:

\begin{equation}\label{eq:GS}
    \begin{split}
    \Delta^* \psi &= -\mu_0 R J_\phi \\
    J_\phi &= J_{\phi}^{pla} + J_{\phi}^{ext} \\
    J_{\phi}^{pla} &= RP'(\psi) + \frac{FF'(\psi)}{\mu_0 R}
    \end{split}
\end{equation}

Here $\psi$ is the poloidal flux, $J_\phi$ is the toroidal current density, $P(\psi)$ is the pressure profile and $F(\psi):= RB_\phi$ is related to the poloidal current. $J_{\phi}^{ext}$ is the current density in the shaping coils and surrounding vacuum vessel structures, and serves as a boundary condition for the free-boundary problem. Analytical solutions are known for only a limited number of cases and in general \cref{eq:GS} must be solved numerically in an iterative process.

\subsection{Data inputs and outputs}

The ground truth training data is composed of equilibria generated from the EFIT01 reconstruction algorithm \cite{Lao1985,Sabbagh2001} during the 2015-2016 NSTX-U campaign. EFIT01 is a magnetics-only equilibrium reconstruction code which is run automatically after each shot and thus offers a sizeable database of samples ($\sim$25000 equilibria, spaced 5-10ms apart, from 220 shots). One limitation of magnetics-only reconstructions is that the pressure and current profiles are not well-constrained. On NSTX-U there does exist a small number of discharges for which kinetic equilibria exist (pressure and current constrained by Thomson and Motional-Stark-Effect diagnostic data). However, EFIT01 is sufficiently accurate for boundary and shape control purposes and its usage is beneficial due to the higher number of samples. It is known that EFIT01 is sufficiently accurate because the NSTX-U shape controller internally uses RTEFIT, and EFIT01 is more accurate than RTEFIT. These two codes are mostly identical but without real-time constraints EFIT01 can run for more iterations until convergence.

\begin{figure}[H]
	\centering 
	\includegraphics[width = 1\columnwidth]{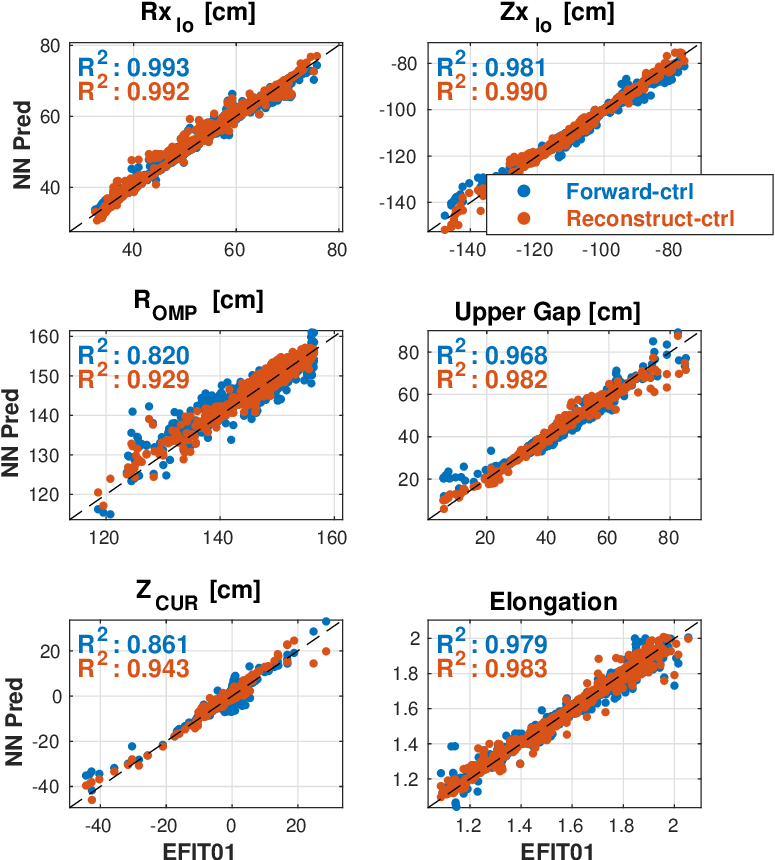}
	\caption{Regression plots of the Eqnet predictions vs EFIT01-measured parameters including: lower x-point position ($Rx_{lo}, Zx_{lo}$), plasma outer midplane radius $R_{OMP}$, upper control segment gap (see \cref{fig:eqnet_pred}), vertical position of the current centroid $Z_{curr}$, and plasma elongation. The reconstruction mode performs slightly better than forward mode as determined by $R^2$ coefficient.}
	\label{fig:regression}
\end{figure}

As mentioned, the data inputs and outputs vary depending on the mode of operation. In the forward mode, the NN maps the coil currents, vacuum vessel currents, P' and FF' profiles to the equilibrium flux map. This mode is useful for simulations and optimizing coil current trajectories, since most of these data inputs are readily available from the shape control model (described in more detail in \cref{sec:response_model}). The shape control model integrates a state vector $\vec{I}$ consisting of the plasma, coil, and vessel currents and thus all the current sources are available as inputs at each time step. Often, additional states $\beta_p$ and $l_i$ are included to track information about the pressure and current ($l_i$ corresponds with the peaked-ness of the plasma current profile). These 2 scalars are almost information-equivalent to knowing the EFIT01 P' and FF' profiles, which by design consist of 1 and 3 basis functions respectively. (We train with the P' and FF' profiles instead of $\beta_p, l_i$ for flexibility in use cases, since \cref{eq:GS} is the standard formulation). 

The output of the model is the equilibrium flux values at all grid locations $\psi_g$, although for better data compression, the NN actually outputs only the plasma flux $\psi_g^{pla}$. The total flux is the sum of the plasma flux and flux from the external sources.

\begin{equation}\label{eq:pla_flux}
    \begin{split}
        \psi_g &= \psi_g^{pla} + \psi_g^{ext} \\
        \psi_g^{ext} &= M_{gc} I_c + M_{gv} I_v
    \end{split}
\end{equation}

Here, $M_{gc}$ and $M_{gv}$ are the mutual inductance matrices between each grid point and the coil and vessel elements respectively, and $I_c$ and $I_v$ are the currents in the coil and vessel elements. The advantage of predicting the plasma flux $\psi_g^{pla}$ is that it has less spatial variation than the total flux leading to better principal component analysis (PCA) data compression and higher accuracy in the final solution. Since the quantities that constitute $\psi_g^{ext}$ are known, there is no disadvantage to only predicting $\psi_g^{pla}$. 

A bit of care must be taken when considering Eqnet's reconstruction version, since in most reconstruction applications the algorithm would only have access to measurements of the currents $I_c$ and $I_v$. This can have a moderate impact: for example, on NSTX the relative error difference between measured and fitted currents was 0.56\% and 4.0\% for coil currents and vessel currents respectively \cite{Sabbagh2001}. When training the reconstruction version, we use the measured values of $I_c$ and $I_v$ in \cref{eq:pla_flux}. In order to still obtain the correct total flux prediction, the NN is trained to predict the plasma flux plus any flux error due to the measurement process. This model is trained using 75 B-probes, 55 flux loops, and 35 vessel current signals and we report strong accuracy in the reconstruction, slightly outperforming the forward mode (\cref{fig:regression}) and RTEFIT (\cref{sec:RTEFIT_comparison}).

The control-oriented modes do not predict the flux but instead predict parameters that are normally derived from the equilibrium. These include: the plasma minor and major radius, elongation, gap sizes of several control segments, positions of the upper and lower x-points, and position of the current centroid. The advantage of this mode is that it bypasses computational steps such locating the touch-point/x-point and tracing the boundary. A shape controller can provide feedback based directly on the NN outputs. A potential downside to this mode is that the outputs are pre-determined, offering less flexibility. If, for example, the position of the shape control segments are adjusted, the NN would have to be retrained with the new control segments. 

\subsection{Data processing}\label{sec:data_processing}

The 220-shot dataset is first organized chronologically by shot number, with an average $\sim$100 samples per shot. The chronologically latest 10\% of shots are set aside for testing. All of the shot numbers and examples referenced in this manuscript are from this test data set. The remaining 90\% of shots are later divided and used for training and validation which is described in more detail in \cref{sec:model_architecture}. This data split procedure is used to mimic the standard experimental application in which a model would be trained on all available data and then need to perform adequately in a future experiment. Random splitting would likely over-estimate the NN capabilities since samples are highly correlated within a shot, and to a lesser-degree, among nearby shots. 

\begin{figure}[h]
	\centering 
	\includegraphics[width = 0.8\columnwidth]{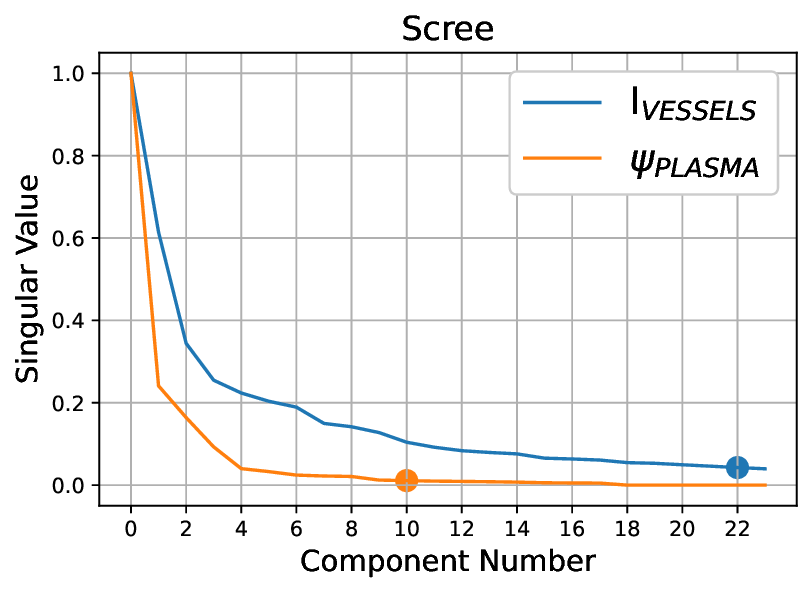}
	\caption{A scree test for 2 representative variables. The number of components is verified to be larger than the ``elbow'' location.}
	\label{fig:scree}
\end{figure}

\begin{figure}[h]
	\centering 
	\includegraphics[width = 1.0\columnwidth]{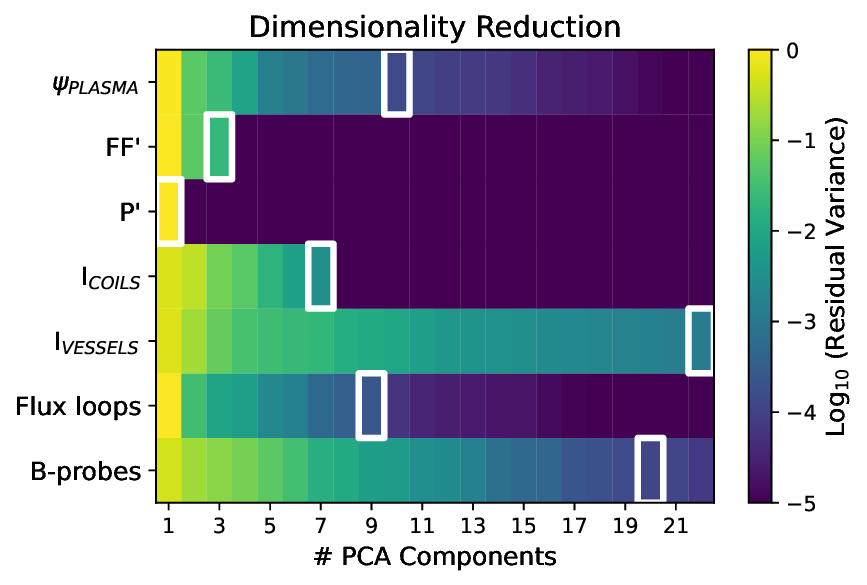}
	\caption{Dimensionality reduction of the Eqnet dataset using Principal Component Analysis (PCA). The outlined boxes indicate 99.5\% captured variance and the number of compenents used for each variable.}
	\label{fig:eqnet_pca}
\end{figure}

Each of the target and predictor variables is dimensionally reduced by performing principal component analysis (PCA) on the training set. The number of components kept is selected to achieve $99.5\%$ captured variance, and is summarized for each variable in \cref{fig:eqnet_pca}. Another rule-of-thumb sometimes used to select the number of components is the scree test shown in \cref{fig:scree} for some representative signals. This test plots the singular values of the data and selects a minimum number of components based on the location of the ``elbow'' where the values begin to level off. We verify that the number of components selected is always greater than this point of maximum curvature.

It should be noted that we do not perform any selection process for including samples as has been done in some previous NN solvers, such as restricting to the flattop portion of the discharge. We include all available samples ranging from t=30ms near the end of breakdown phase and concluding at shot termination (end of rampdown, disruption, and/or loss of control). Indeed, a sizeable fraction of shots terminate off-normal ($>30\%$ terminate within $0.6\si{s}$, while the target pulse length is generally $1-2\si{s}$) which is a significant contributor to the range and variance in the data, and important for learning.

In order to adequately learn non-flattop equilibria, one step that proved to be necessary for good performance was performing a PCA “merging” procedure on the target variable $\psi_g^{pla}$. This was done by separating the dataset into rampup, flattop, and rampdown/pre-disruption phases of the discharge, performing PCA on each of these phases separately, and combining the PCA components together to form a single basis. The motivation is that the flattop phase tends to be over-represented because Ip is constant and the flux surfaces and plasma position are mostly constant. In contrast, much of the important variation occurs during rampup and rampdown/pre-disruption. Naively using PCA on the combined data did not represent the rampup or rampdown equilibria well, as determined by visual inspection of the projected flux surfaces. This problem is not fixed by oversampling the rampup and rampdown equilibria, likely because these samples have lower Ip and therefore lower magnitude of $\psi_g^{pla}$ so they still do not sufficiently contribute to the variance in the data matrix.

The PCA-merging step is performed as follows: suppose we have a data matrix $X$ of dimensions $n_{features} \times n_{samples}$ with sample mean $\mu$, and that we have performed PCA on $X$ and have a set of principal components $U$ and coefficients $W$. We recall that the data matrix can be projected from the components and coefficients as

\begin{equation}\label{eq:pca}
    X = UW + \mu
\end{equation}

Thus for two (truncated) PCA projections $X_1 = U_1 W_1 + \mu_1$ and $X_2 = U_2 W_2 + \mu_2$, the combined projection $[X_1 \; X_2]$ is within the span of the combined components and means $\bar U := [U_1 \; U_2 \; \mu_1 \; \mu_2]$. Thus we can write 

\begin{equation}
    \left[ X_1 \;\; X_2 \right] = \underbrace{\left[ U_1 \;\; U_2 \;\; \mu_1 \;\; \mu_2\right]}_{\bar U} \bar W 
\end{equation}

for some set of weights $\bar W$, or a version with nonzero mean which is in the same PCA-like form as \cref{eq:pca}. 

\begin{equation}\label{eq:merge_pca}
    \begin{split}
        \left[ X_1 \;\; X_2 \right] &= \underbrace{\left[ U_1 \;\; U_2 \;\; \mu_1 \;\; \mu_2\right]}_{\bar U} \bar W + \mu^* \\
        \mu^* &= (\mu_1 + \mu_2) / 2
    \end{split}
\end{equation}

\Cref{eq:merge_pca} is not a valid PCA projection because the columns of $\bar U$ are not orthogonal, but we can obtain a set of orthogonal components $U^*$ by performing, for example, a singular value decomposition of $\bar U$. The merged coefficients are then $W^* = {U^*}^T([X_1 \;\; X_2] - \mu^*)$. This simple preprocessing step proved to be sufficient for compressing data across a wide range of samples. The final result compresses the $\psi_g^{pla}$ grid of size $[65 \times 65]$ to  just 10 PCA coefficients.  

\Cref{fig:eqnet_pca} summarizes data compression results for the various inputs and outputs. For the flux values, significant dimensionality reduction is achieved. As mentioned, the P' and FF' profiles consist of 1 and 3 basis functions respectively, which is captured by this mode decomposition. The highest dimensional inputs are related to the vessel currents and B-probe diagnostics. After dimensionality reduction, we normalize the inputs to have zero mean and unit variance, to arrive at the neural network inputs. The standardization and PCA decompositions are then inverted to map neural network outputs to physical outputs. 

\begin{figure*}[h]
	\centering 
	\includegraphics[width = 18cm]{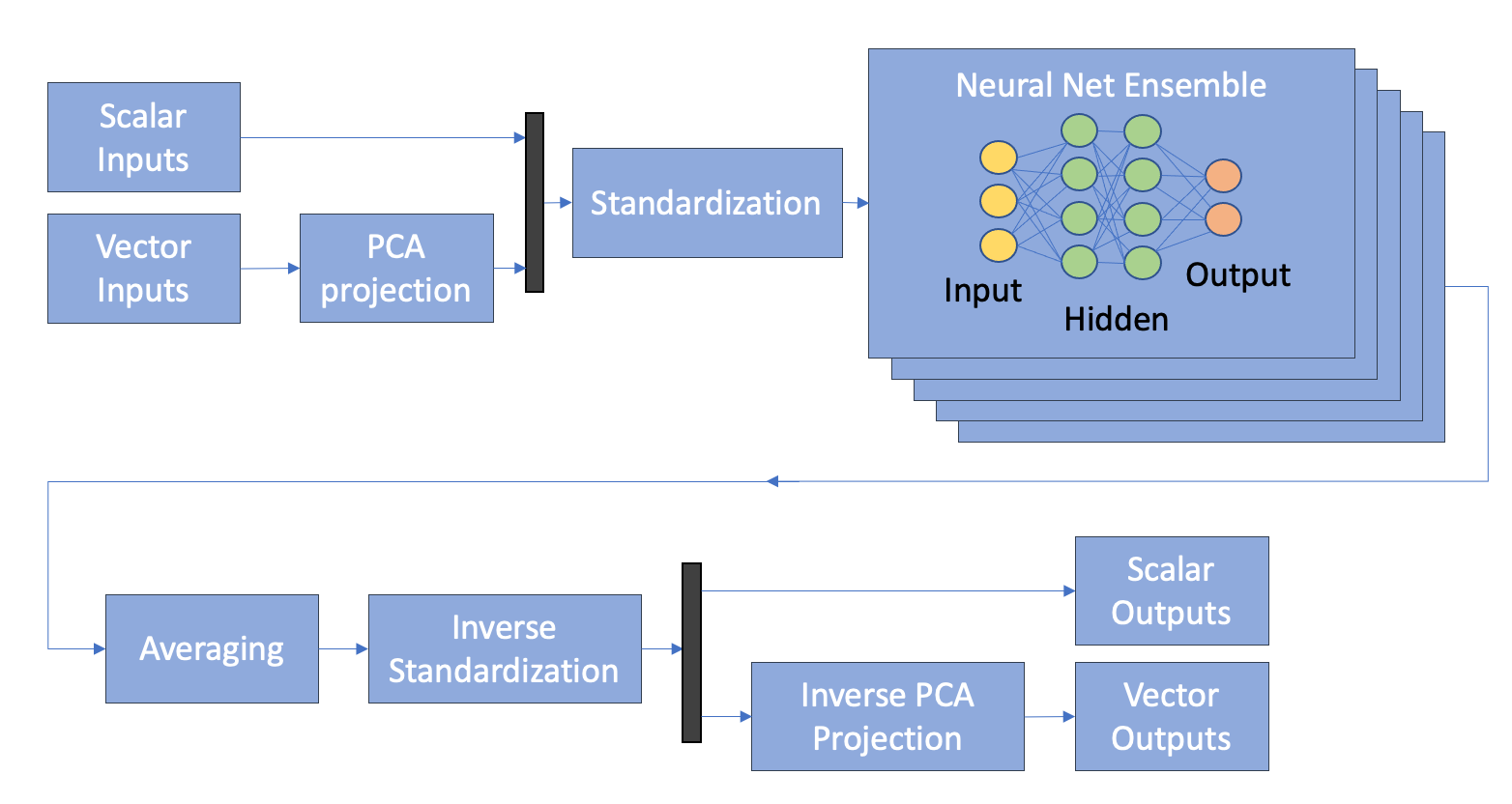}
	\caption{Model archictecture for Eqnet and Pertnet including data processing steps. The core of the model is an ensemble of fully-connected networks.}
	\label{fig:architecture}
\end{figure*}

\subsection{Model architecture and hyperparameter tuning}\label{sec:model_architecture}

The model architecture including data processing steps is shown in \cref{fig:architecture}. The core of the model uses fully-connected artificial neural nets (ANNs) which are created and trained in PyTorch with the Adam optimizer. We use a small (5-member) ensemble of NNs and average their predictions, before inverse-transforming the results back to the original data-space. Recall that 10\% of shots are reserved for testing with the remaining 90\% shots going to training and validation. Each NN ensemble member is trained on a random subset (90\%) of shots from within this training/validation set (i.e. an individual NN will see 90\% of the 90\%). This ensemble method is used for two reasons--first, to increase the robustness of the final prediction, since the various applications tend to need stable predictions, and second, as a method for determining statistical properties of the predictions. (This ensemble method is similar to 5-fold cross-validation, with each ensemble member representing one iteration of the cross-validation procedure.) As will be shown in \cref{sec:uncertainty}, the agreement between ensemble members can be a useful tool for gauging uncertainty in the prediction, and actually does correlate with underlying uncertainty in the equilibrium reconstruction.

\begin{figure*}[h]
	\centering 
	\includegraphics[width = 18cm]{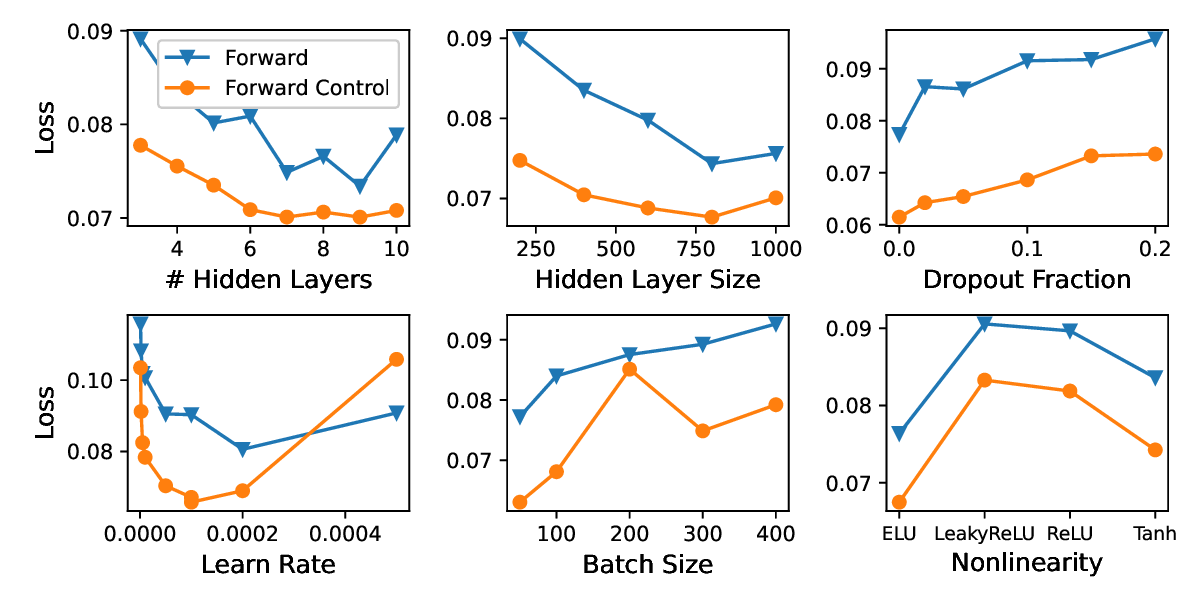}
	\caption{Hyperparameter grid scans used to identify the best performing Eqnet model. The loss value is computed using the validation dataset. The landscape is fairly convex making it straightforward to identify good parameter combinations.}
	\label{fig:eqnet_hyperparams}
\end{figure*}

The network hyperparameters are chosen by performing scans of the number of hidden layers, hidden layer size, dropout fraction, optimizer learn rate, batch size, and nonlinearity activation function. Results of this procedure are shown in \cref{fig:eqnet_hyperparams}. The hyperparameter landscape is fairly convex, making it straightforward to identify a suitable parameter combination. We note that the NN tends to perform better with larger network sizes (depth and breadth), zero dropout, and small batch sizes. We also note a jump in performance when using the ELU activation function as opposed to the standard ReLU.

\subsection{Performance}

\subsubsection{Model timing}
A key motivation for this work is the computational speed of neural networks. The time to train an individual NN for 300 epochs is roughly 8-10 hours on PPPL's portal cluster (single core AMD Opteron 6136 2.4GHz CPU). More important is the inference time which was measured on a Macbook Pro 3.5GHz Dual-Core Intel i7 CPU and ranges from 810$\mu$s to 940$\mu$s depending on the particular NN version. This was computed in Python and represents the inference time for a single NN, since it would be natural to parallelize the inference of each NN ensemble member. This is already fast enough to support real-time operation (approximately 1ms, which is also approximately the time needed by RTEFIT to complete a cycle), although additional speed increases are expected for when the model is translated into a compiled programming language. Therefore we expect any real-time applications could be supported without issue.

\subsubsection{General performance}

\begin{figure*}[h]
	\centering 
	\includegraphics[width = 18cm]{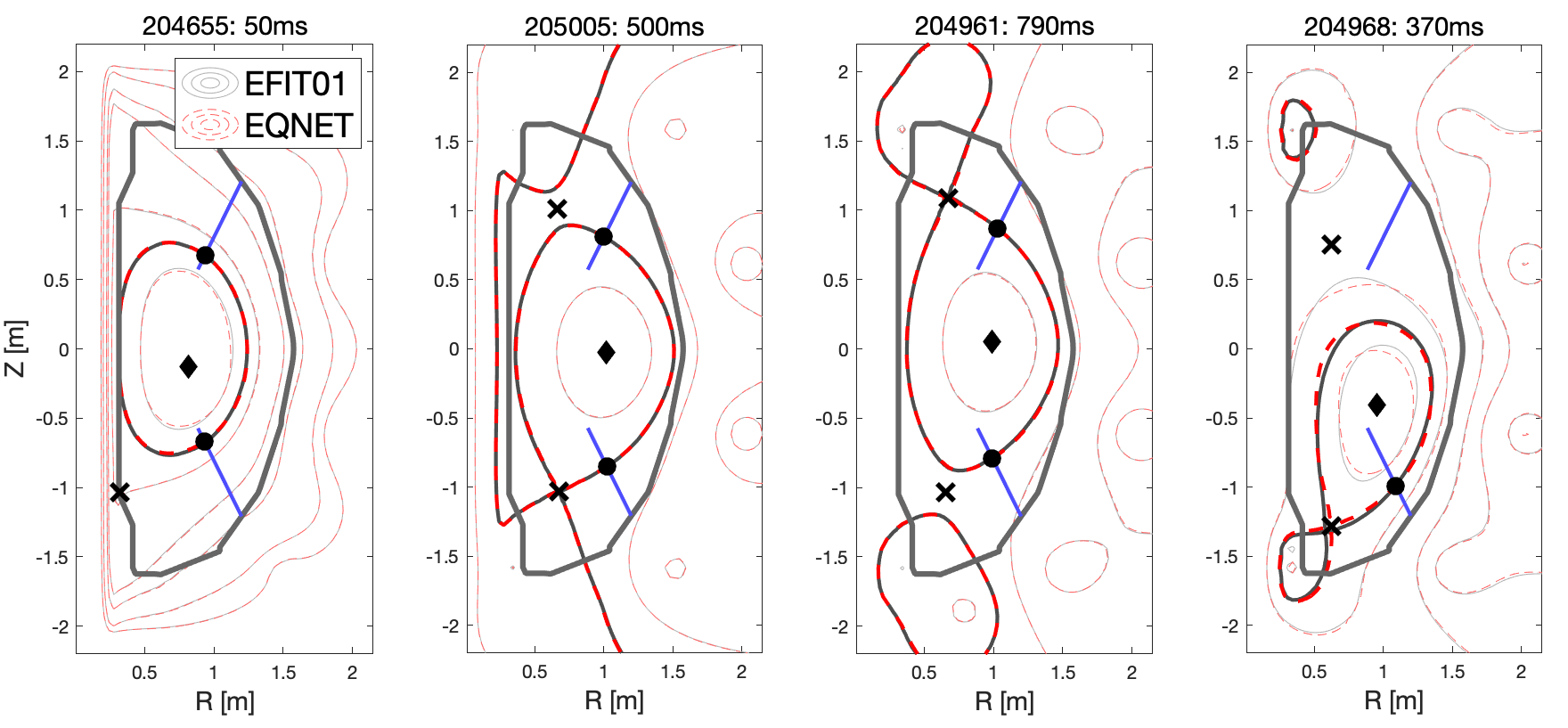}
	\caption{Equilibrium flux surfaces for several test shots as determined by EFIT01 and the Eqnet neural network. The boundary contour is shown in bold. Also shown are control segments (blue), magnetic axis (black diamond), and x-points.}. 
	\label{fig:eqnet_pred}
\end{figure*}

\Cref{fig:eqnet_pred} shows some representative results for NN flux surface predictions (forward-mode) taken from the test dataset. Also depicted are the x-points, magnetic axis, and control segments used in computing the upper and lower gap. The 4 samples shown represent different categories of equilibrium shapes such as limited plasma produced during rampup, lower single-null (LSN) during flattop, upper single-null (USN) during flattop, and a LSN from a disrupting plasma. In the first three cases the flux surfaces are mostly indistinguishable by eye at this scale. One does observe errors in the disruption sample. Since this type of event is off-normal it is less well-represented in the training data. This suggests that for less-represented samples or shapes that have not yet been explored on NSTX-U (e.g. negative triangularity), there is room for future improvement by obtaining more experimental data or supplementing with synthetic data. 

A regression test of shaping parameters is shown in \cref{fig:regression} for the control-oriented NNs. We observe that the reconstruction mode outperforms the forward mode slightly in most variables as measured by the $R^2$ coefficient. ($R^2=1$ is perfect prediction) A possible explanation is that the reconstruction mode has significantly more NN inputs to learn from, including highly localized information from the diagnostics. The hardest variables to predict are the outer midplane radius and current centroid position with $R^2$ values 0.820/0.929 and 0.861/0.943 respectively. As another quantitative measure of accuracy we report absolute prediction error values at the 50th (median) and 90th percentile levels in \cref{table:percentiles}. For most geometric features the median errors are generally only a few mm, and the 90th percentile errors are about 1cm. 

\begin{table}[h]
	\centering 
	\includegraphics[width = 1.0\columnwidth]{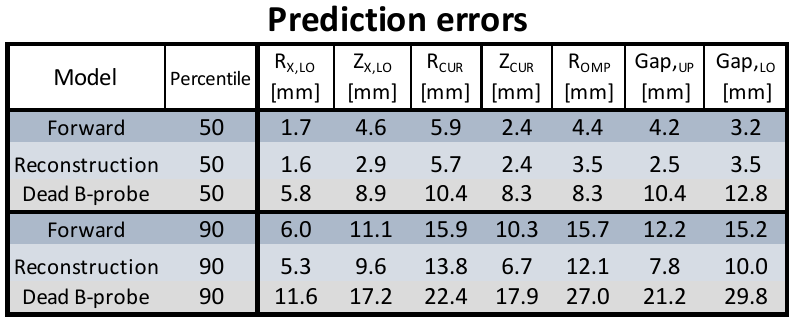}
	\caption{Prediction errors for several models computed at the 50th and 90th percentiles. Most errors are less than 1cm. Zeroing out a magnetic probe results in an approximate doubling of errors.}
	\label{table:percentiles}
\end{table}

While the large majority of samples are predicted accurately, it is true that some samples have errors of multiple cm which is high for shape control. The time trace predictions shown in \cref{fig:eq_timetraces} give some qualifying insight. From the $R_{OMP}$ and $Z_{CUR}$ plots we observe multiple oscillations in the signals, which tend to occur very early in the shot (rapidly changing conditions during current rampup) or near the end of a shot (from disruption, loss of control, or rapidly varying conditions during rampdown). In fact, for errors $\geq3$cm, $68\%$ of bad NN predictions occur during the first or last 100ms of the shot, and 100\% within the first or last 200ms. These dynamic features are representative of the NSTX-U campaign and is in part due to the control difficulties of associated with a short-pulse, highly-elongated, unstable plasma. The worst errors always occur during these fast transient events, and represents real physical uncertainty since on this fast time scale induced vessel currents do not have time to decay, and have a shielding effect on the diagnostics. In other words the underlying EFIT01 equilibria also have more uncertainty at these times. 
 
The timetraces in \cref{fig:eq_timetraces} includes predictions from the forward version, where shaping parameters have been derived from the flux surfaces, and from the forward-control version (direct prediction of the parameters). Both sets of NN predictions match each other and EFIT01 closely. In these figures, the control version has slightly less error overall but this is not universal. Neither version is substantially superior to the other so the main tradeoff is in terms of the desired output type, with flux surface prediction being the more flexible option.

\subsubsection{Ensemble agreement as an indicator of prediction uncertainty}\label{sec:uncertainty}

\begin{figure*}[h]
	\centering 
	\includegraphics[width = 16cm]{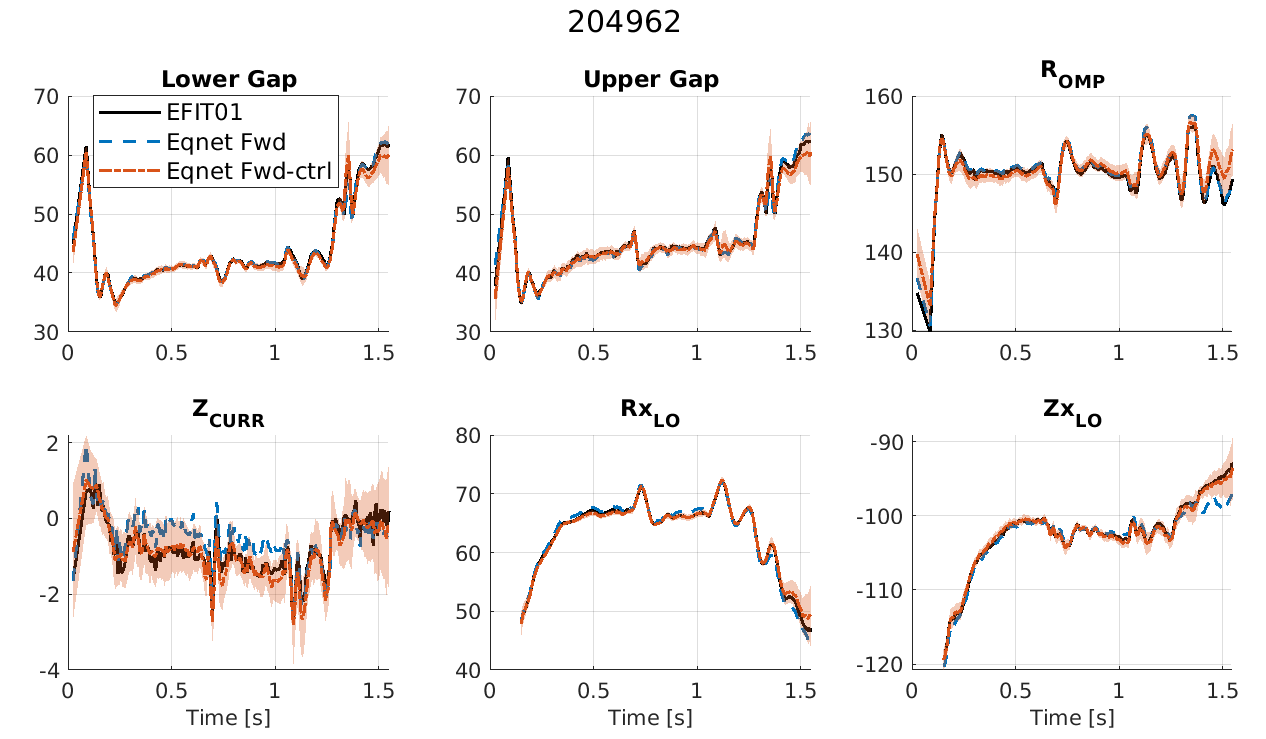}
	\caption{Time traces of several geometric quantities for test dataset shot 204125, as measured by EFIT01 and predicted by Eqnet. The lower and upper gaps are the intersection of the plasma boundary with the control segments as shown in \cref{fig:eqnet_pred}. The shaded red region represents $\pm$3 standard deviations between the predictions of NN ensemble members.}
	\label{fig:eq_timetraces}
\end{figure*}

We advocate that the disagreement between ensemble members can be used as an indicator for prediction uncertainty. The logic is that since the NNs were trained with different shot data and initialization weights, they will tend to give different predictions for samples that are hard to predict. This is not a direct conversion; often the difference between NNs is smaller than the difference of the NN vs EFIT01. The shaded red areas in \cref{fig:eq_timetraces} is a measure of agreement between the underlying NN ensemble members in Eqnet, plotted at $\pm$3 standard deviations. We observe the disagreement between NNs is not constant but varies throughout the shot, for example, increasing substantially for all signals in the last 300ms of the shot. This is also the time period where prediction errors are larger, and where it is known that there is more uncertainty in the reconstruction (due to the fast dynamics at beginning or end of shot). This information could be used to help with safe deployment of NN-based applications, for example, by relying on Eqnet only if the disagreement is below a certain threshold.

\subsubsection{Comparison vs RTEFIT}\label{sec:RTEFIT_comparison}

As discussed in the introduction, one of the possible applications for Eqnet is to perform real-time equilibrium reconstruction during shots. On NSTX-U this task is performed by the RTEFIT algorithm. The reconstruction version of Eqnet can functionally act like RTEFIT since it is trained to predict equilibria using the same magnetic signals as inputs. Additionally, Eqnet is trained on the EFIT01 outputs which is a more accurate reconstruction than RTEFIT, so there is a possibility for Eqnet to outperform RTEFIT by some standards. 

To clarify the relationship, EFIT01 is the offline magnetics-only reconstruction and RTEFIT is the online real-time version. There are a few key differences between these codes. Most importantly, EFIT01 performs more iterations and achieves better convergence of the Grad-Shafranov solution, whereas RTEFIT only has time to compute one iteration per equilibrium. Second, EFIT01 can perform non-causal filtering of the diagnostics since it has access to future measurements for applying, e.g., drift correction. Lastly, RTEFIT is only available starting at 100-150ms into the shot whereas EFIT01 is run offline for times larger than $\sim30$ms. 

One source of error for RTEFIT stems from occasional failure to reach convergence. This failure is more likely to occur during large changes to the plasma, such as strong oscillations, or if the boundary-defining point is switching between various locations. Another source of error is related to the vessel eddy currents. As noted in the original NSTX implementation, the vessel current measurements are computed via Ohm's law $I = V_{meas} / R$ neglecting dynamical effects \cite{Gates2004,Sabbagh2001}, which is accurate enough to prevent ``an unphysical solution, yet is large enough to permit the solution to deviate from the computed value during periods  of fast current transients.'' \cite{Sabbagh2001} Vessel currents can also have the effect of temporarily shielding magnetic sensors from changes that are happening to the plasma, negatively impacting the reconstruction. 

\begin{figure}[h!]
	\centering 
	\includegraphics[width = 1.0\columnwidth]{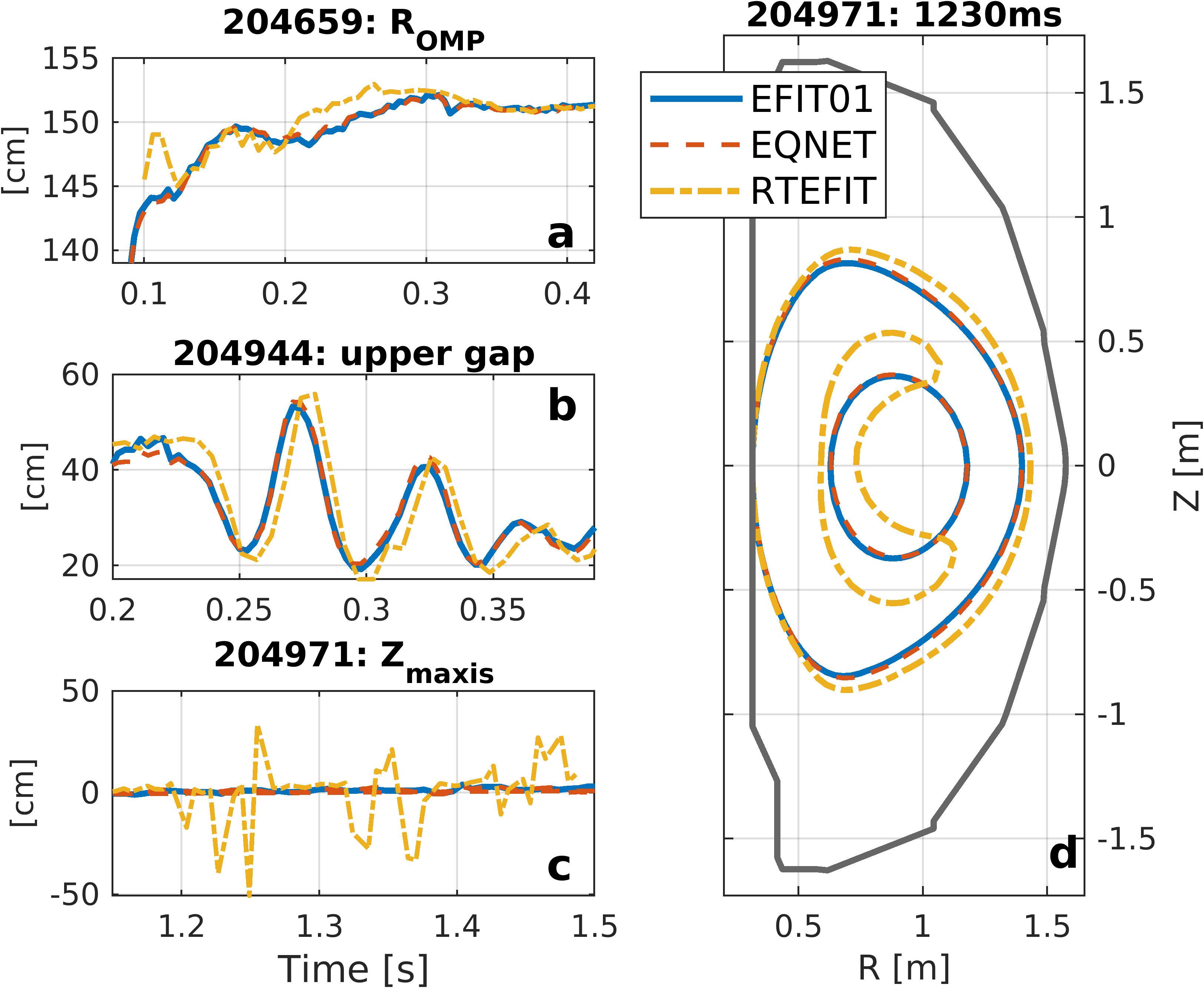}
	\caption{Comparison of Eqnet and RTEFIT. The RTEFIT error and 5ms phase shift in \textbf{(a)} and \textbf{(b)} is due to dynamic events inducing strong vessel eddy currents. Eqnet has learned to compensate for this. The RTEFIT mismatch in \textbf{(c)} and \textbf{(d)} is due to insufficient convergence.}
	\label{fig:RTEFIT_comparison}
\end{figure}

\Cref{fig:RTEFIT_comparison} compares several reconstructed signals from EFIT01, Eqnet, and RTEFIT for a few shots from the test dataset, and highlights that Eqnet can outperform RTEFIT especially in relation to the two error sources (convergence and induced vessel currents). 

The first panel \ref{fig:RTEFIT_comparison}a depicts $R_{OMP}$ the radius of the plasma boundary at the outer midplane. Eqnet matches EFIT01 closely, while RTEFIT deviates from EFIT01 at $t=0.10-0.13$ and again at $t=0.20-0.29s$. The former period is due to RTEFIT being switched on and requiring several iterations to converge. The latter error is due to the vessel current effect. During this time $0.20-0.29s$ the plasma was changing from limited to diverted and moving radially outward. This radial plasma movement induces strong vessel currents, and RTEFIT does not coincide with EFIT01 until this motion stops and eddy currents die down. 

In \ref{fig:RTEFIT_comparison}b we observe the measurements of the upper gap during a shot that was experiencing undesired vertical oscillations. The RTEFIT signal is phase-shifted behind EFIT01 by $5$ms indicating the length of time the vessel dynamics are exerting strong influence. By contrast Eqnet is in-phase with EFIT01 demonstrating that the neural net has learned how to compensate for these dynamic effects. 

Convergence-related issues are shown in \ref{fig:RTEFIT_comparison}c and \ref{fig:RTEFIT_comparison}d. During the time period shown in \ref{fig:RTEFIT_comparison}c, the plasma was oscillating significantly, and the boundary point was switching rapidly between an inner-wall touch point, upper x-point, and outer-wall touch point. This led to insufficient convergence for RTEFIT, and so $Z_{maxis}$, the vertical position of the magnetic axis swings wildly by $>50$cm. (Note that RTEFIT will aggressively move the magnetic axis, but the boundary measurements are not changing by quite this magnitude). One of these equilibria at $t=1.23$s is plotted in \ref{fig:RTEFIT_comparison}d. While the boundary is relatively close to the EFIT01 boundary, it is clear that the internal flux contours are very different. 

These benefits of a NN-based approach must be weighed against disadvantages such as a weaker ability to generalize, and the need for high-quality training data that may not exist for, e.g. a newly built machine. While a NN may not be immediately usable during new machine commissioning, it could increasingly offer benefits as a reconstruction backup as data is collected over time. 

\subsubsection{Robustness to diagnostic failure}
Another aspect to consider for deployment is robustness of the NN to sensor failure. For example, if one of the magnetic probes dies, it would be beneficial if experiments could continue as normal with only a slight decrease in performance. To investigate this issue, we trained the reconstruction-version with all sensor inputs, and then zeroed out one of the magnetic probes and measured the predictions. This was repeated for all 75 magnetic probes, and the average of these results is recorded in \cref{table:percentiles}. We note that most errors have roughly doubled compared to the original reconstruction-mode predictions. For many of the signals, the 90th percentile errors are now in the range of 2-3cm. This would likely be workable, but with reduced operational capability. For example, the $R_{OMP}$ target would likely have to decrease to give margin for wall clearance. These results were obtained using light (1\%) dropout during training to randomly zero out some magnetic sensor inputs. It is likely possible to improve these results further by optimizing the dropout rate, to strike a balance between peak performance and robustness. 

\subsection{Design Application}\label{sec:design} 

As a simple application task we attempt to solve an inverse equilibrium design problem using the Eqnet forward-version. The design problem is: given a target shape, what are the coil currents that will transform an equilibrium into that shape. Notationally, we have target shape $\vec{y}$ and the neural net is a map from inputs $\vec{x}$ to an equilibrium $f(\vec{x})$. (We restrict to only modifying coil currents, not all of the network inputs.) The design procedure is carried out by using the Matlab function \textbf{fminsearch}, which is a standard optimization algorithm based on the Nelder-Mead method. At each iteration, the search makes a call to the NN (which evaluates $f(\vec{x})$) and adjusts the inputs to minimize the cost function $J := (f(\vec{x}) - \vec{y})^2$. 

\begin{figure}[h]
	\centering 
	\includegraphics[width = 1.0\columnwidth]{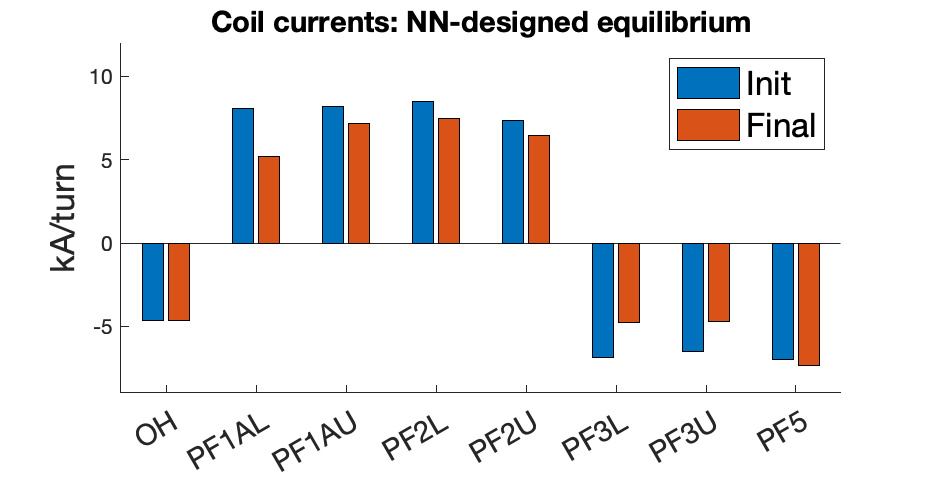}
	\caption{Coil currents designed by with fminsearch + Eqnet in order to construct an elongated equilibrium.}
	\label{fig:design_coils}
\end{figure}

\begin{figure}[h]
	\centering 
	\includegraphics[width = 1.0\columnwidth]{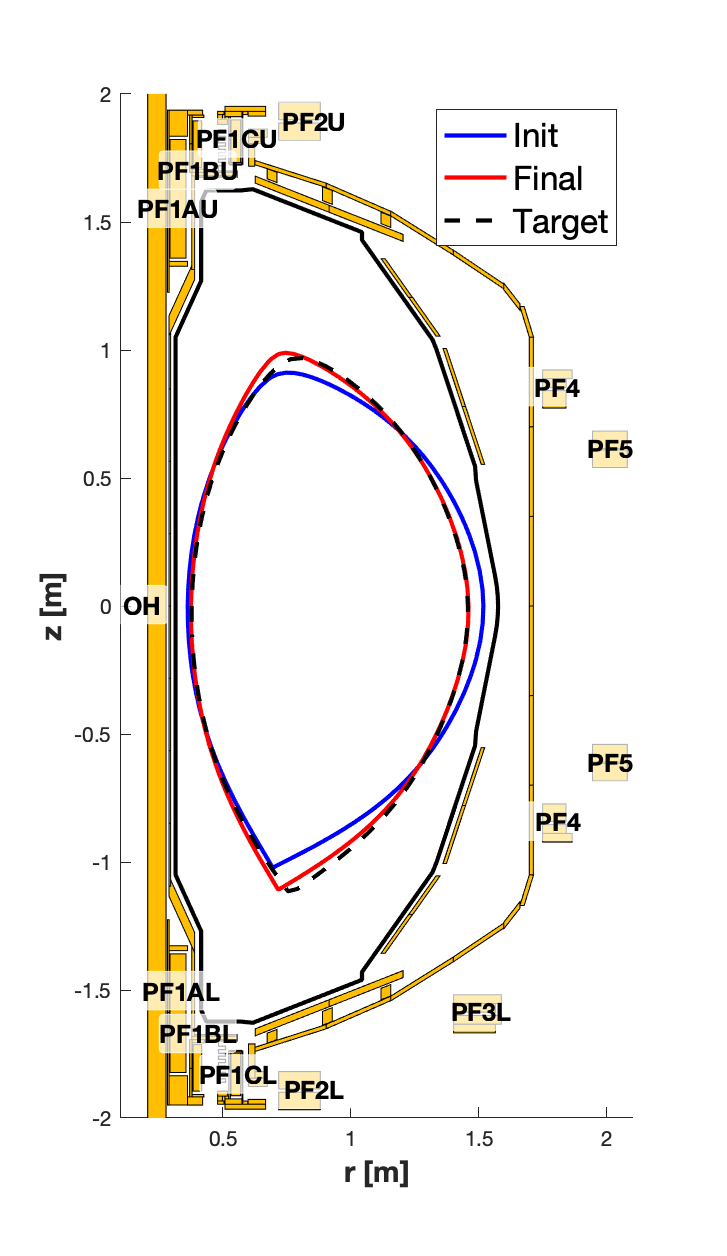}
	\caption{Equilibrium designed with fminsearch + Eqnet, by adjusting the coil currents in order to match the target boundary. The target shape is an elongated $\kappa=1.95$ LSN equilibrium.}
	\label{fig:design_eq}
\end{figure}

The target shape is obtained by beginning with a LSN diverted equilibrium and increasing the elongation from 1.65 to 1.95. The target boundary also has a smaller minor radius 54cm vs 58cm and larger inner and outer gaps. The initial, target, and final shapes are shown in \cref{fig:design_eq}, and coil currents in \cref{fig:design_coils}. Note that the boundary ``final'' boundary is not the NN-prediction (which exactly matches the target), but the result of a standard free-boundary calculation using the NN-designed coil currents. The equilibrium produced by the NN is indeed much closer to the target, with some small differences in the boundary at the lower and upper. Given that this is a novel equilibrium for the NN, which has not seen this exact equilibrium before and was not generated via EFIT, we expect this gives a reasonable estimate of generalization error. This level is also acceptable for feedforward planning applications, since feedback control can correct for small errors. 

\section{Pertnet}\label{sec:pertnet}

\subsection{Plasma response model}\label{sec:response_model}

``Pertnet'' is a neural network we have designed to predict the nonrigid plasma response, which is a term related to the Grad-Shafranov equation that arises in the development of the shape control model. This section includes some background context on the shape control model and plasma response. For further detail on shape control, we refer readers to the review paper by Ambrosino and Albanese \cite{Ambrosino2005}. The nonrigid plasma response describes how the plasma current shifts and redistributes in response to changes in the currents of surrounding conductors. The response can be written in units of current or units of flux which are related by the grid mutual inductance:

\begin{equation}\label{eq:pla_response}
    \pdv{\psi_g^{pla}}{I_i} = M_{gg} \pdv{I_g^{pla}}{I_i}
\end{equation}

Here, $\psi_g^{pla}$ is the grid flux due to plasma currents, $I_g^{pla}$ is the plama current distribution, and $M_{gg}$ is the mutual inductance matrix for the grid. We use the ``pla'' superscript to distinguish flux that is due to plasma sources, since the total flux at any point is due to both plasma and external sources. The physical interpretation is that when the current in an external conductor is perturbed, this changes the background magnetic field. Since the plasma is carrying current it experiences a Lorentz force that causes it to shift back into equilibrium. We rely on the Gspert code \cite{Welander2005} to obtain the plasma response, which is based on a linearization to the Grad-Shafranov equation. In addition to obtaining the response due to current changes, it is often useful to obtain the response due to changes in global parameters such as $\beta_p$ and $l_i$. These are included in the Pertnet predictions, although not explicitly discussed here. 

The plasma response arises in several places in the shape control model. For example, the model often begins by writing a circuit equation for all the conducting elements within the tokamak:

\begin{equation}\label{eq:circuit}
    \begin{split}
        v_i &= R_i I_i + \dot \psi_i \\
            &= R_{i} I_i + M_{ij} \dot I_j + \pdv{\psi_i^{pla}}{I_j} \dot I_j \\
            &= R_i I_i + M_{ij} \dot I_j + M_{ig} \pdv{I_g^{pla}}{I_j} \dot I_j
    \end{split}
\end{equation}

where $v_i$ is the applied voltage from external power supplies, $R_i$ is the resistance of each element. We see that the plasma response contributes to the flux change at conductors and must be accounted for when modeling current evolution. Additionally, this term is needed for modelling how various feedback parameters (x-points, current centroid) change with respect to the inputs. Consider some inputs $x$ that define the equilibrium ($x$ includes the coil currents), and some shaping parameters $y$ that are derived from the equilibrium flux. This relation is 

\begin{equation}
    y = f(\psi(x))
\end{equation}

$f(\cdot)$ is the mapping from equilibrium to parameters, and $\psi(\cdot)$ is the map from inputs to equilibrium. By the chain rule, in order to see how the inputs affect the shape parameters we need access to the plasma response.

\begin{equation}\label{eq:chainrule}
    \pdv{y}{x} = \pdv{f}{\psi_g}\left( \pdv{\psi_g^{ext}}{x} + \pdv{\psi_g^{pla}}{x} \right)
\end{equation}

We train 2 versions of the Pertnet model -- a standard flux mode that calculates the flux response $\pdv{\psi_g^{pla}}{x}$ and a control-oriented mode that directly estimates the parameter responses $\pdv{y}{x}$. To illustrate this mapping from flux response to parameter response via \cref{eq:chainrule}, we show the derivation for 3 parameters (current centroid response, x-point response, and vertical growth rate) which are the main parameters predicted by the Pertnet control mode:

\subsubsection{Current centroid response}
The current centroid position is the current-weighted average of each coordinate over the plasma domain $\Omega$:

\begin{equation}
    Z_{cur} = \frac{1}{I_p}\sum_\Omega{Z_g I_g} 
\end{equation}

Applying \cref{eq:chainrule} and \cref{eq:pla_response} gives

\begin{equation}\label{eq:cur_response}
    \pdv{Z_{cur}}{x} = \frac{1}{I_p}\sum_\Omega{Z_g \left( M_{gg}^{-1}\pdv{\psi_{g,pla}}{x} \right)}
\end{equation}

\subsubsection{X-point response}

The x-point is the location at which the flux gradient is equal to zero, so finding the x-point is equivalent to finding the root of the equation  $F(\vec{s}) = [\psi_r(\vec{s}) \; \psi_z(\vec{s})]^{T}$. Using Newton's method with the Jacobian of $F$ defined $J_F$, one could find the root by  successively applying updates as

\begin{equation}
    \Delta \vec{s} = -J_F(\vec{s})^{-1}F(\vec{s})
\end{equation}   

If we have already converged to an x-point, but then apply a perturbation to the flux, the response of the x-point is 

\begin{equation}\label{eq:xp_response}
    \pdv{ (r_x, z_x)}{x} = \pdv{\Delta \vec{s}}{x} = -J_F(\vec{s})^{-1} \begin{bmatrix} \pdv{\psi_r}{x} \\[0.6em] \pdv{\psi_z}{x} \end{bmatrix}
\end{equation}

which depends explicitly on the plasma response via the last terms since

\begin{equation}
    \pdv{\psi_r}{x} = \pdv{}{r} \left( \pdv{\psi^{ext}}{x} + \pdv{\psi^{pla}}{x} \right)
\end{equation}

\subsubsection{Vertical growth rate}
The vertical growth rate is an instability that can be related to the evolution of currents in conducting structures. Defining the response term in \cref{eq:circuit} as $\pdv{\psi_i^{pla}}{I_j} := X_{ij}$, this equation can be rewritten in matrix form as

\begin{equation}
    \vec v = \vec R \vec I + (\vec M + \vec X) \dot{\vec{I}}
\end{equation}

or equivalently

\begin{equation}\label{eq:statespace}
\begin{split}
    \dot{\vec{I}} &= \vec A \vec I + \vec B \vec v \\
    \vec A &:= -(\vec M + \vec X)^{-1} \vec R \\
    \vec B &:= (\vec M + \vec X)^{-1} \\
\end{split}
\end{equation}

which is the standard form of the shape control state space model. The vertical growth rate is taken as the largest unstable eigenvalue of the $\vec{A}$ matrix. 

\begin{equation}\label{eq:gamma}
    \gamma = \text{max(real(eig(} \textbf{A} \text{)))}
\end{equation}

\begin{figure}[h]
	\centering 
	\includegraphics[width = 1.0\columnwidth]{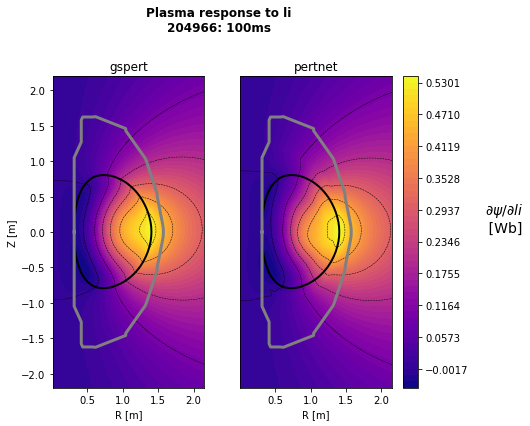}
	\caption{Comparison of the plasma response to an $l_i$ perturbation as calculated by the Gspert code and pertnet neural network approximator. The plasma boundary is overlayed in black.}
	\label{fig:response204069_100}
\end{figure}

\begin{figure}[h]
	\centering 
	\includegraphics[width = 1.0\columnwidth]{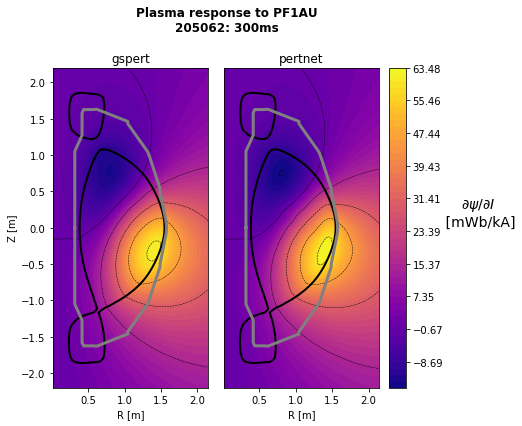}
	\caption{Comparison of the plasma response to a PF1AU current perturbation as calculated by the Gspert code and pertnet neural network approximator. The plasma boundary is overlayed in black.}
	\label{fig:response204069_500}
\end{figure}

\begin{figure}[h]
	\centering 
	\includegraphics[width = 1.0\columnwidth]{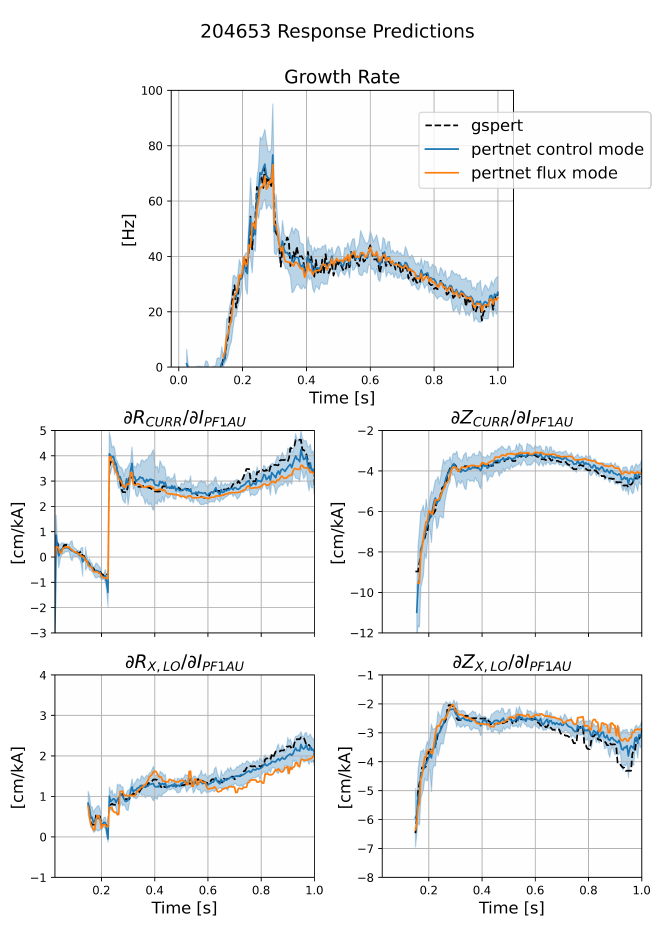}
	\caption{Predictions of the plasma response for several parameters (growth rate, current centroid response to PF1AU, and x-point position response to PF1AU) as calculated by Gspert and versions of Pertnet. The Pertnet control mode directly predicts these parameters, while the pertnet flux mode predicts the flux response which maps to the parameter responses. The shaded blue region corresponds to $\pm$4 standard deviations of the disagreement between NN ensemble members.}
	\label{fig:pertnet_timetraces}
\end{figure}

\begin{figure}[h]
	\centering 
	\includegraphics[width = 1.0\columnwidth]{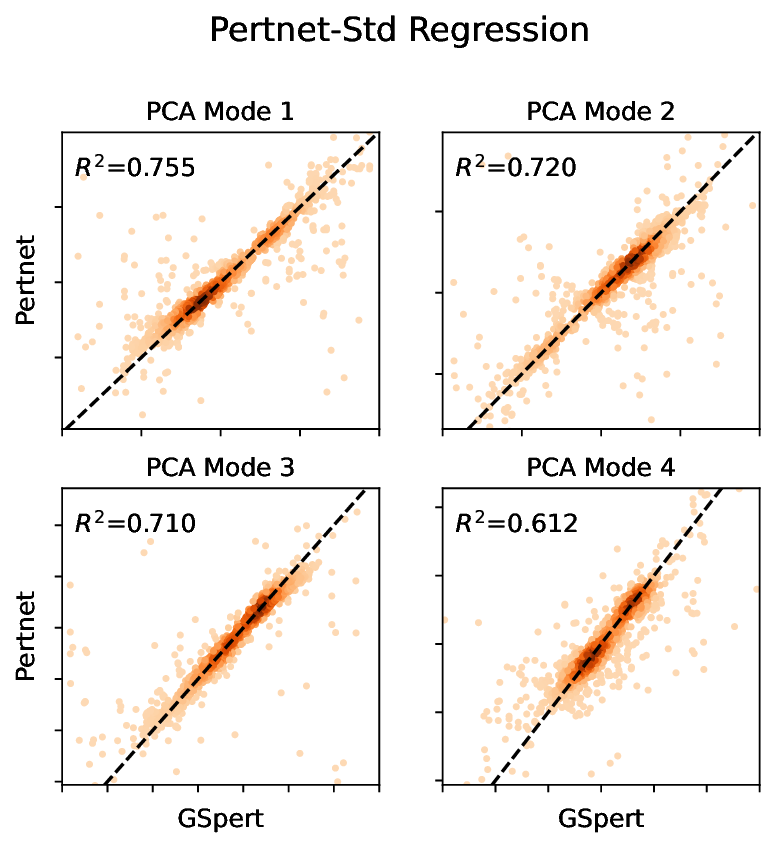}
	\caption{Regression of the Pertnet flux prediction vs calculated values from Gspert. The data shown is the test dataset for the response to PF5. Each plot shows the coefficient of the corresponding PCA mode of the flux response.}
	\label{fig:pertnet_std_regression}
\end{figure}

\begin{figure}[h]
	\centering 
	\includegraphics[width = 1.0\columnwidth]{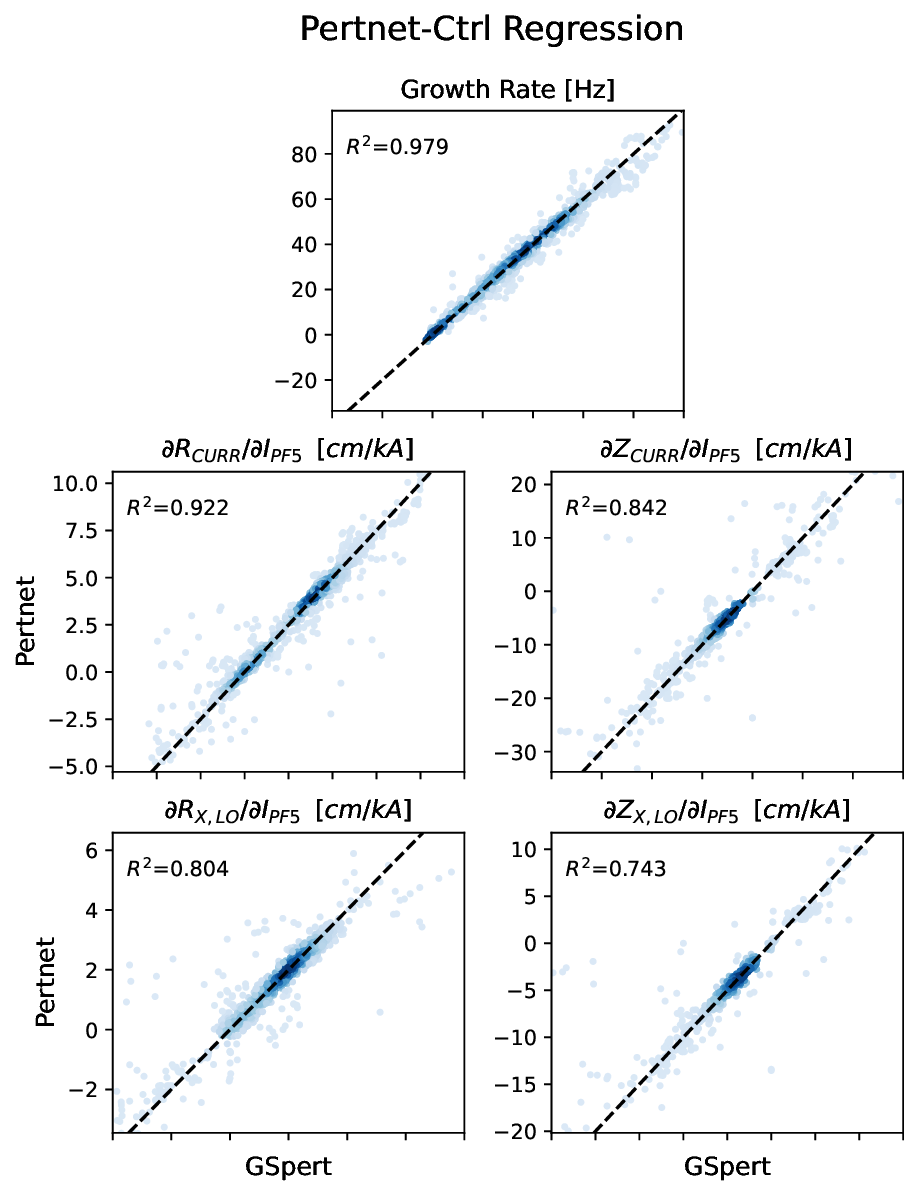}
	\caption{Regression of Pertnet parameter responses vs the Gspert values, based on predictions from Pertnet-control mode.}
	\label{fig:pertnet_ctrl_regression}
\end{figure}

\subsection{Data Inputs and Outputs}
Similar to Eqnet, we train 2 versions of the Pertnet model -- a ``standard'' flux mode that calculates the flux response $\pdv{\psi_g^{pla}}{x}$ and a ``control'' mode that predicts the parameter responses $\pdv{y}{x}$. The standard mode is more general in that all aspects of the shape control model can be derived from it, albeit with more computational effort. On the other hand, control mode skips some computational steps in finding the shaping parameter response directly. This would make it useful for experiments in providing fast estimates to the shape controller. A potential downside is that if the desired outputs are modified then the NN requires retraining. Additionally, the control mode does not predict all aspects of the shape control model, such as what the full $\vec{X}$ matrix is. 

For both modes, the inputs are a complete description of the equilibrium. We assume more inputs here than we did for Eqnet because we are linearizing around the current equilibrium and can thus assume access to it, either via a NN prediction or reconstruction algorithm. The inputs include: $\psi(r,z)$ as calculated by EFIT01, the EFIT01 pressure $P'(\psi)$ and $FF'(\psi)$ profiles, current distribution $J_\phi$, coil and vessel currents, $I_p$, coordinates of the boundary contour, $\psi$ at the magnetic axis and boundary, and the $(R,Z)$ position of the magnetic axis, x-points, and current centroid. 

For the standard mode, the output is the flux response on the grid. Note that from \cref{eq:pla_response} we could predict in units of flux or units of currents. For the NN, it is advantageous to predict in flux units. This is because the current response can contain highly localized information about the current variation in nearby grid cells. By contrast the flux response has less spatial variation and has better data compression. In other words, the grid mutual inductance $M_{gg}$ acts like a smoothing or averaging filter when transforming from current to flux. 

\subsection{Architecture and hyperparameter tuning}

We use the same set of $\sim$25000 equilibria samples and chronological train-validation-test data splitting method used in Eqnet (\cref{sec:eqnet}). For each sample, we run Gspert to find the plasma response to the 8 active PF coils, 40 vessel circuits, $I_p$, $\beta_p$, and $l_i$. For the standard flux mode, we train an individual NN to predict the response of each quantity. This performs better than predicting the flux response to all inputs simultaneously, which does not dimensionally reduce very well, since each response quantity can vary significantly in magnitude. For the control mode, which predicts responses of individual parameters that are of much lower dimension than the flux response, it was found that a single NN was sufficient. 

\begin{figure}[h]
	\centering 
	\includegraphics[width = 1.0\columnwidth]{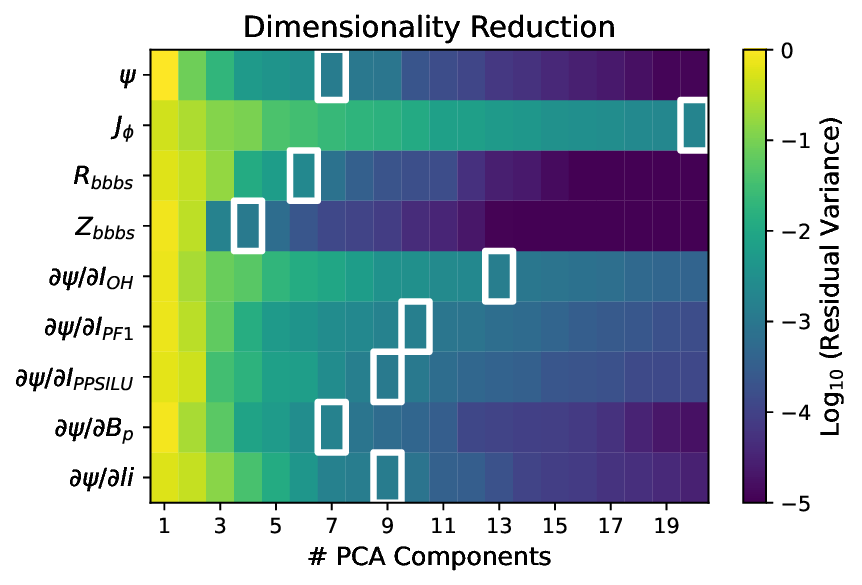}
	\caption{Dimensionality reduction of the Pertnet dataset using Principal Component Analysis (PCA). The outlined boxes indicate 99.5\% captured variance and the number of compenents used for each variable.}
	\label{fig:pertnet_pca}
\end{figure}

Principal component analysis is used to compress all the inputs and outputs, choosing the number of components to capture $99.5\%$ variance, with results of this procedure shown in \cref{fig:pertnet_pca}. Only some of the response quantities are plotted. Note that the equilibrium currents $J_\phi$ has a much larger representation than the flux $\psi$, again indicating that the flux is generally smoother than the current. All response variables have the similar order of magnitude ($\sim$10 components). Some variables, such as the ohmic (OH) coil have expectedly higher variation. The OH coil spans the entire vertical length of NSTX-U, and thus can interact with the plasma from multiple angles depending on plasma position. After PCA, the coefficients are then normalized to zero mean and unit variance. The full architecture of the model is identical to Eqnet and shown in \cref{fig:architecture}.

\begin{figure*}[h]
	\centering 
	\includegraphics[width = 18cm]{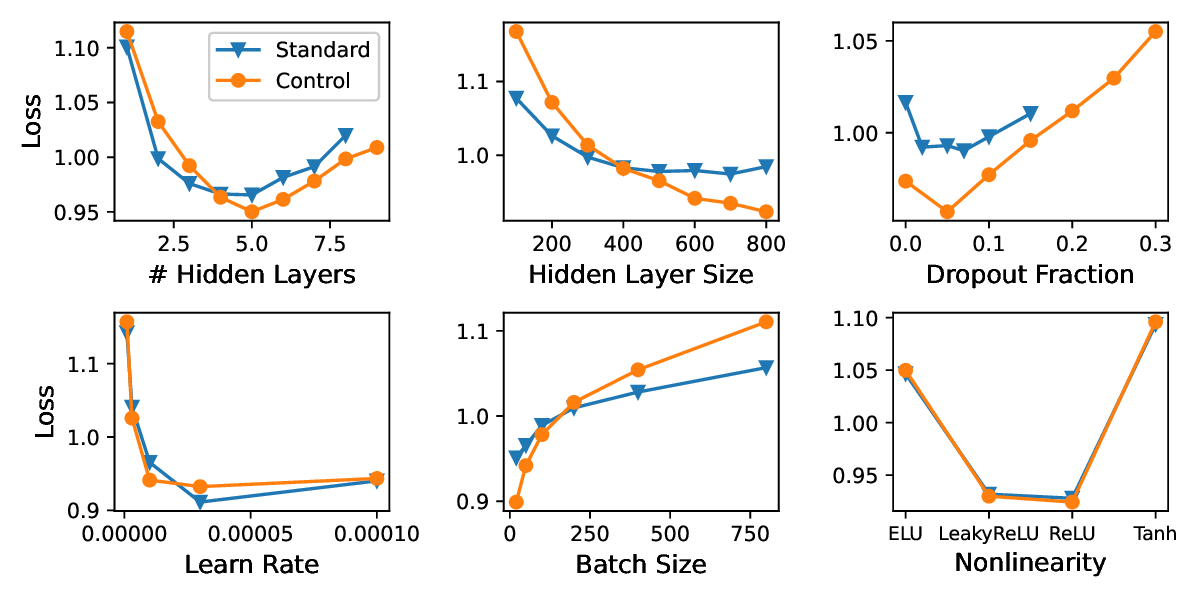}
	\caption{Hyperparameter grid scans used to identify the best performing Pertnet model. The optimal model depends on which Pertnet mode is under consideration, with the control mode favoring larger networks than the standard mode.} 
	\label{fig:pertnet_hyperparams}
\end{figure*}

Training is performed in Pytorch with the Adam optimizer. Hyperparameters are selected by performing grid scans over the various model parameters such as size and depth, with results shown in \cref{fig:pertnet_hyperparams}.  Early on, it was observed visually that using the L1 loss (absolute error) function performed better than the standard L2 loss (sum of squares). This is theorized to be a result of noise and outliers in the dataset. In particular, a small fraction of samples were observed to have spurious results due to running Gspert automatically on all 25k+ equilibria. These include errors such as not converging fully to features of interest, or not identifying the boundary correctly in a situation with multiple x-points and touch points. An attempt was made to perform automated filtering of the outputs, but manual inspection of all results was not performed. In addition, the samples are theorized to have a lot of noise due the inherent noise associated with taking derivatives. Since L2 penalizes squared errors, the network was probably learning too much from predictions with large errors that should have been identified as outliers. Another possible indicator of stochasticity is that the performance improves by including $5-10\%$ dropout on the network weights during training. The final network is smaller than Eqnet with the standard/control modes having 4/5 layers with hidden layer size 400/800. Note that the hyperparameters differ significantly between the modes because they are actually problems with relatively different structure, since the standard mode has a NN for each flux response, and the control mode is a single NN that predicts all parameter responses. 

\subsection{Performance}

\subsubsection{Model timing}
Using the same setup as before (Macbook 3.5GHz Dual-Core Intel i7 CPU), the inference times for individual NNs are measured at 230$\mu$s (standard flux mode) and 710$\mu$s (control mode). This indicates these NNs could be adapted to real-time applications without issue. Since the flux mode is composed of many NNs (one for each response variable) this may require a parallelized implementation in order to keep computational times low.

\subsubsection{General prediction performance}
Performance of the flux response prediction is shown in \cref{fig:pertnet_std_regression} where we have plotted the NN-predicted PCA coefficient vs the Gspert values, for the first 4 PCA modes. These predictions are for the flux response to PF5 the midplane outboard shaping, for the entire test dataset. The densest clusters lie close to the true values, though there is scatter for individual sample points. As expected, the prediction is most accurate on the first PCA mode ($R^2=0.755$) and decreases slightly with each increasing mode number. 

\Cref{fig:response204069_100} and \cref{fig:response204069_500} show individual flux responses to two different perturbation sources, $l_i$ and the upper inboard coil PF1AU. The predicted vs actual flux responses  are difficult to distinguish by eye, although there are some slight differences. 
Predictions of the timetraces for parameter responses are plotted in \cref{fig:pertnet_timetraces} for as single shot 204653. These include the current centroid response, x-point response, and growth rate as obtained by \cref{eq:cur_response,eq:xp_response,eq:gamma}. Even though the flux response is not trained to predict these parameters, the extracted parameter response is close to the Gspert values. However, if only interested in the parameter response, the control mode outperforms slightly. For example, in the second half of the shot, Gspert estimates that the radial position of the x-point ($R_{X,LO}$) will grow increasingly more sensitive to the PF1AU coil. The control-mode captures this trend, though the standard flux mode does not trend as strongly. The growth rate prediction for both models matches Gspert very accurately, which is a nice confirmation of prediction of all the NN models. Note that the growth rate prediction is unlike the other parameters, in that it does not depend on the response to just one input coil, but all the coils simultaneously (via \cref{eq:statespace,eq:gamma}, since the $\vec{X}$ matrix includes effects from all coils). Even if individual flux responses can each have inaccuracies, this is an indicator that the dominant effects upon the coil current dynamics is being captured. 

Lastly, \cref{fig:pertnet_ctrl_regression} shows the performance of Pertnet-control mode for predicting individual parameter responses, in this case to PF5. We note that the individual responses generally have slightly higher correlation coefficients than prediction of the flux modes. While it is not a linear transformation between flux and parameters, this is consistent with the observation on \cref{fig:pertnet_timetraces} that direct parameter prediction performed slightly better.

The most accurate prediction is the growth rate ($R^2$ = 0.979) while the the least accurate is the Z-derivative of the lower x-point ($R^2$ = 0.743). We record that 50th and 90th percentile errors for these respective parameters are $1.2/4.6$Hz and $0.16/1.00 \si{[cm/kA]}$. This is a fairly large difference between the 50th and 90th percentiles. Again, this is a result of the largest responses and prediction errors being driven by a small number of samples as a result of the dynamic periods during rampup and loss-of-control (e.g. for the worst 5\% of predictions, 88\% occur within the first or last 200ms of the shot). This tends to manifest itself worse in the vertical direction than in the radial direction, consistent with the vertical instability effect. Note the scale difference for R-coordinate response vs the Z-coordinate responses, and that the $R^2$ coefficient for radial responses is higher indicating better predictions for this direction.

\section{Conclusion}

In this work we have introduced two neural nets for the NSTX-U tokamak, as part of a suite of prediction and simulation tools being developed for optimizing plasma scenarios. The neural nets include Eqnet, a reconstruction or forward free-boundary equilibrium solver, and Pertnet which calculates the nonrigid plasma response. As a simple application, we use Eqnet to design the coil currents for a desired target equilibria, indicating good performance and stability. Both neural nets are trained and tested with different combinations of input and output data. We find that using Eqnet to reconstruct equilibria from magnetic diagnostics gives some performance improvements over RTEFIT especially when vessel eddy currents are significant. We report strong performance for Pertnet in predicting the plasma response. If only interested in the response of various parameters, it is simpler (fewer NNs to train) and marginally more accurate to predict these directly. However, predicting the full flux response is needed in some situations and can give good estimates of the parameters with additional computational steps.


\section*{Acknowledgements}
Thanks to the anonymous reviewers for many suggestions that strengthened the content of this manuscript. This work was supported by the US Department of Energy Grant under contract numbers DE-AC02-09CH11466, DE-SC0015480, DE-SC0021275, and DE-SC0015878.

\bibliography{bib}

\end{document}